\documentstyle[aps,12pt]{revtex}
\widetext
\draft
\tighten
\begin{document}

\preprint{EFUAZ FT-96-39-REV}

\title{On the importance of the
normalization\thanks{Submitted
to ``Journal of Physics A".}}

\author{{\bf Valeri V. Dvoeglazov}}

\address {Escuela de F\'{\i}sica, Universidad Aut\'onoma de Zacatecas \\
Apartado Postal C-580, Zacatecas 98068, ZAC., M\'exico\\
Internet address:  valeri@cantera.reduaz.mx\\
URL: http://cantera.reduaz.mx/\~~valeri/valeri.htm}

\date{First version: December 1996, Final version: November 1997}

\maketitle

\baselineskip13pt

\medskip

\begin{abstract}
We repeat the known procedure of the derivation of
the set of Proca equations. It is shown that it can be written
in various forms.  The importance of the normalization  is point out for
the problem of the correct description of  spin-1 quantized fields.
Finally, the discussion of the so-called Kalb-Ramond field is presented.
\end{abstract}

\newpage

Recently, a new concept of the longitudinal magnetic field, the ${\bf
B}^{(3)}$ field, was proposed~\cite{EV1}. It is based on the equation
containing the cross product of two transversal modes of electromagnetism:
\begin{equation}
{\bf B}^{(1)} \times {\bf B}^{(2)}
= i B^{(0)} {\bf B}^{(3)\,\ast}\quad,
\end{equation}
and represents itself a non-trivial generalization of the Maxwell's
electromagnetic theory (see ref.~\cite{DVO96R} for the list of other
generalizations). This concept advocates the mass of photon
and appears to be in the contradiction with the concept of the
$m\rightarrow 0$ group contraction for a photon
as presented by Wigner and Inonu~\cite{WI} and Kim~\cite{Kim}.

On the other hand, in~\cite{DVO96} it was shown that the angular
momentum generators $J_{\kappa\tau}$ are equated to zero after the
application of the generalized Lorentz condition $\partial_\mu F^{\mu\nu}
=0$ (and dual to that) to the quantum states,  when considering the theory
of quantized antisymmetric tensor fields (the representation $(1,0)\oplus
(0,1)$ of the Lorentz group, according to the ordinary wisdom).  The
formula (9) of the cited reference tells us:\footnote{While this formula
is the consequence of the particular choice of the Lagrangian of
antisymmetric tensor field, nevertheless, the main conclusion is hold for
other types of Lagrangians, including those which possess the Kalb-Ramond
gauge invariance~\cite{Hayashi}.}
\begin{eqnarray}
J_{\kappa\tau} &=& \int
d^3 {\bf x} \left [ (\partial_\mu F^{\mu\nu}) (g_{0\kappa} F_{\nu\tau} -
g_{0\tau} F_{\nu\kappa}) - (\partial_\mu F^\mu_{\,\,\,\,\kappa}) F_{0\tau}
+ (\partial_\mu F^\mu_{\,\,\,\,\tau}) F_{0\kappa} + \right. \nonumber\\
&+& \left.  F^\mu_{\,\,\,\,\kappa} ( \partial_0 F_{\tau\mu} + \partial_\mu
F_{0\tau} +\partial_\tau F_{\mu 0}) -   F^\mu_{\,\,\,\,\tau} ( \partial_0
F_{\kappa\mu} +\partial_\mu F_{0\kappa} +\partial_\kappa F_{\mu 0}) \right
]\quad. \label{gene}
\end{eqnarray}
Therefore, the spin vector reads\footnote{Presenting this formula we,
however, do {\it not} think that the question of the correct definition of
the relativistic spin vector is so simple as believed before.}
\begin{equation}
{\bf J}^k = {1\over 2} \epsilon^{ijk} J^{ij} =
\epsilon^{ijk} \int d^3 {\bf x} \left [ F^{0i} (\partial_\mu F^{\mu j}) +
F_{\mu}^{\quad j} ( \partial^0 F^{\mu i} + \partial^\mu F^{i0} +\partial^i
F^{0\mu}) \right ]\quad.
\end{equation}
Thus, the helicity of the ``photon"
described by the physical fields ${\bf E}$ and ${\bf B}$ is
believed~\cite{Hayashi,Avdeev} to be equated to zero (?), what is in the
strong contradiction with experimental results.\footnote{M. Kalb and P.
Ramond claimed explicitly~[6c, p. 2283, the third line from below]:
``thus, the massless $\phi_{\mu\nu}$ has one degree of freedom".
While they call $\phi_{\mu\nu}$ as ``potentials" for the field
$F^{\alpha\beta\gamma} = \partial^\alpha \phi^{\beta\gamma}
+\partial^\beta \phi^{\gamma\alpha} +\partial^\gamma \phi^{\alpha\beta}$,
nevertheless, the physical content of the antisymmetric tesnor
field of the second rank (the representation $(1,0)\oplus (0,1)$ of the
Lorentz group) must be in accordance with the requirements of relativistic
invariance.} ``The helicity -- the projection of the spin onto the
direction of motion -- proves to be equal to zero \ldots even without the
restriction to plane waves, the 3-vector of spin [formula (12) of the
cited paper] vanishes on solutions \ldots", ref.~[7b].  Moreover, if this
construct describes the $h=0$ fields it seems to contain internal
theoretical inconsistencies because is in the contradiction with the
Weinberg theorem $B-A = h$, ref.~\cite{WEIN0}.  This argument was also
used earlier by Evans~\cite{EV1,EV2,EV3}.   In order to understand the
nature of these contradictions one should reveal relations between the
Evans-Vigier concept and the concept of the so-called Kalb-Ramond
field~\cite{Hayashi}.  While some insights into the problem have been
already presented~\cite{EV3,DVO96} in this paper we try to deepen the
understanding of apparent conflicts, which were mentioned.

We believe in the power of the group-theoretical methods in the analyses
of the physical behaviour of different-type classical (and quantum)
fields. We also believe that the Dirac equation can be applied to
some particular quantum states of the spin $1/2$. Finally, we believe that
the spin-0 and spin-1 particles can be constructed by taking the direct
product of the spin-1/2 field functions~\cite{BW}. So, on the basis of
these postulates let us firstly repeat the Bargmann-Wigner procedure of
obtaining the equations for bosons of spin 0 and 1.
The set of basic equations for  $j=0$ and $j=1$ are written, e.g.,
ref.~\cite{Lurie}
\begin{mathletters} \begin{eqnarray} \left [
i\gamma^\mu \partial_\mu -m \right ]_{\alpha\beta} \Psi_{\beta\gamma} (x)
&=& 0\quad,\label{bw1}\\ \left [ i\gamma^\mu \partial_\mu -m \right
]_{\gamma\beta} \Psi_{\alpha\beta} (x) &=& 0\quad.\label{bw2}
\end{eqnarray} \end{mathletters}
We expand the $4\times 4$ matrix wave function into the antisymmetric and
symmetric parts
\begin{mathletters}
\begin{eqnarray}
\Psi_{[\alpha\beta ]} &=& R_{\alpha\beta} \phi +
\gamma^5_{\alpha\delta} R_{\delta\beta} \widetilde \phi +
\gamma^5_{\alpha\delta} \gamma^\mu_{\delta\tau} R_{\tau\beta} \widetilde
A_\mu \quad,\label{as}\\ \Psi_{\left \{ \alpha\beta \right \}} &=&
\gamma^\mu_{\alpha\delta} R_{\delta\beta} A_\mu
+\sigma^{\mu\nu}_{\alpha\delta} R_{\delta\beta} F_{\mu\nu}\quad,
\label{si}
\end{eqnarray} \end{mathletters}
where $R= CP$ has the
properties (which are necessary to make expansions (\ref{as},\ref{si}) to
be possible in such a form)
\begin{mathletters} \begin{eqnarray}
&& R^T = -R\quad,\quad R^\dagger =R = R^{-1}\quad,\\
&& R^{-1} \gamma^5 R = (\gamma^5)^T\quad,\\
&& R^{-1} \gamma^\mu R = -(\gamma^\mu)^T\quad,\\
&& R^{-1} \sigma^{\mu\nu} R = - (\sigma^{\mu\nu})^T\quad.
\end{eqnarray}
\end{mathletters}
The  explicit form of this matrix can be chosen:
\begin{equation}
R=\pmatrix{i\Theta & 0\cr 0&-i\Theta\cr}\quad,\quad \Theta = -i\sigma_2 =
\pmatrix{0&-1\cr 1&0\cr},
\end{equation} provided that $\gamma^\mu$
matrices are in the Weyl representation.  The equations
(\ref{bw1},\ref{bw2}) lead to the Kemmer set of the $j=0$ equations:
\begin{mathletters}
\begin{eqnarray}
m\phi &=&0 \quad,\\
m\widetilde \phi &=& -i\partial_\mu \widetilde A^\mu\quad,\\
m\widetilde A^\mu &=& -i\partial^\mu \widetilde \phi\quad,
\end{eqnarray}
\end{mathletters}
and to the Proca set of the equations for the $j=1$ case:\footnote{We
could use another symmetric matrix $\gamma^5 \sigma^{\mu\nu} R$ in the
expansion of the symmetric spinor of the second rank.
In this case the equations would read
\begin{mathletters}
\begin{eqnarray}
&& i\partial_\alpha \widetilde F^{\alpha\mu} +{m\over 2} B^\mu = 0\quad,
\label{de1}\\
&& 2im \widetilde F^{\mu\nu} = \partial^\mu B^\nu -\partial^\nu
B^\mu\quad.\label{de2}
\end{eqnarray} \end{mathletters}
in which  the dual tensor
$\widetilde F^{\mu\nu}= {1\over 2} \epsilon^{\mu\nu\rho\sigma}
F_{\rho\sigma}$ presents,
because we used that in the Weyl representation
$\gamma^5 \sigma^{\mu\nu} = {i\over 2} \epsilon^{\mu\nu\rho\sigma}
\sigma_{\rho\sigma}$; $B^\mu$ is the corresponding vector potential.  The
equation for the antisymmetric tensor field (which can be obtained from
this set) does not change its form (cf.~\cite{DVO96R}) but we see some
``renormalization" of the field functions. In general, it is permitted to
choose various relative phase factors in the expansion of the
symmetric wave function (5b). We would have additional phase factors in
equations relating the physical fields and the 4-vector potentials.  They
can be absorbed by the redefinition of the potentials/fields (the choice
of normalization). The above discussion shows that the dual tensor
of the second rank can also be epxanded in potentials, as opposed to the
opinion of the referee (JPA) of my previous
paper.}$^{,}$\footnote{Recently, after completing this work the
paper~\cite{Kirch} was brought to our attention. It deals with the
redundant components in the $j=3/2$ spin case. If the claims of that paper
are correct we would have to change a verbal terminology which we use to
describe the above equations.}
\begin{mathletters} \begin{eqnarray} &&\partial_\alpha
F^{\alpha\mu} + {m\over 2} A^\mu = 0 \quad,\label{1} \\ &&2 m F^{\mu\nu} =
\partial^\mu A^\nu - \partial^\nu A^\mu \quad, \label{2}
\end{eqnarray}
\end{mathletters}
In the meantime, in the textbooks, the latter set is
usually written as ({\it e.g.}, ref.~\cite[p.135]{Itzyk})
\begin{mathletters}
\begin{eqnarray}
&&\partial_\alpha F^{\alpha\mu} + m^2 A^\mu = 0 \quad,
\label{3}\\
&& F^{\mu\nu} = \partial^\mu A^\nu - \partial^\nu A^\mu \quad,
\label{4}
\end{eqnarray} \end{mathletters} The set (\ref{3},\ref{4}) is
obtained from (\ref{1},\ref{2}) after the normalization change $A_\mu
\rightarrow 2m A_\mu$ or $F_{\mu\nu} \rightarrow {1\over 2m} F_{\mu\nu}$.
Of course, one can investigate other sets of equations with different
normalization of the $F_{\mu\nu}$ and $A_\mu$ fields. Are all these
sets of equations equivalent?  As we shall see, to answer this question
is not trivial. Papers~\cite{DVA} argued that
the physical normalization is such that in the massless-limit
zero-momentum field functions should vanish in the momentum
representation (there are no massless particles at rest). Next, we
advocate the following approach:  the massless limit can and must be taken
in the end of all calculations only, {\it i.~e.}, for physical quantities.

Let us proceed further. In order to be able to answer the question about
the behaviour of the spin operator
${\bf J}^i = {1\over 2} \epsilon^{ijk}
J^{jk}$ in the massless limit
one should know the behaviour of the fields $F_{\mu\nu}$ and/or $A_\mu$ in
the massless limit.  We want to analyze the first set (\ref{1},\ref{2}).
If one advocates the following definitions~\cite[p.209]{Wein}
\begin{mathletters}
\begin{eqnarray}
\epsilon^\mu  ({\bf 0}, +1) = - {1\over \sqrt{2}}
\pmatrix{0\cr 1\cr i \cr 0\cr}\quad,\quad
\epsilon^\mu  ({\bf 0}, 0) =
\pmatrix{0\cr 0\cr 0 \cr 1\cr}\quad,\quad
\epsilon^\mu  ({\bf 0}, -1) = {1\over \sqrt{2}}
\pmatrix{0\cr 1\cr -i \cr 0\cr}\quad, \quad
\end{eqnarray}
\end{mathletters}
and ($\widehat p_i = p_i /\mid {\bf p} \mid$,\, $\gamma
= E_p/m$), ref.~\cite[p.68]{Wein} or ref.~\cite[p.108]{Novozh},
\begin{mathletters}
\begin{eqnarray} &&
\epsilon^\mu ({\bf p}, \sigma) =
L^{\mu}_{\quad\nu} ({\bf p}) \epsilon^\nu ({\bf 0},\sigma)\quad,\\ &&
L^0_{\quad 0} ({\bf p}) = \gamma\, ,\quad L^i_{\quad 0} ({\bf p}) =
L^0_{\quad i} ({\bf p}) = \widehat p_i \sqrt{\gamma^2 -1}\, ,\quad
L^i_{\quad k} ({\bf p}) = \delta_{ik} +(\gamma -1) \widehat p_i \widehat
p_k \quad \end{eqnarray} \end{mathletters}
for the field operator of the 4-vector
potential, ref.~\cite[p.109]{Novozh} or
ref.~\cite[p.129]{Itzyk}\footnote{Remember that the invariant integral
measure over the Minkowski space for physical particles is $$\int d^4 p
\delta (p^2 -m^2)\equiv \int {d^3  {\bf p} \over 2E_p}\quad,\quad E_p =
\sqrt{{\bf p}^2 +m^2}\quad.$$ Therefore, we use the field operator as in
(\ref{fo}). The coefficient $(2\pi)^3$ can be considered at this stage as
chosen for the convenience.  In ref.~\cite{Wein} the factor $1/(2E_p)$ was
absorbed in creation/annihilation operators and instead of the field
operator (\ref{fo}) the operator was used in which the $\epsilon^\mu
({\bf p}, \sigma)$ functions for a massive spin-1 particle were
substituted by $u^\mu ({\bf p}, \sigma) = (2E_p)^{-1/2} \epsilon^\mu ({\bf
p}, \sigma)$, what leads to the confusions in the definitions of the
massless limit $m\rightarrow 0$ for  classical polarization vectors.}$^,
$\footnote{In the
general case it might be useful to consider front-form helicities (and/or
``time-like" polarizations) too.  But, we leave the presentation of
rigorous theory of this type for subsequent publications.}
\begin{equation} A^\mu (x^\mu) =
\sum_{\sigma=0,\pm 1} \int {d^3 {\bf p} \over (2\pi)^3 2E_p} \left
[\epsilon^\mu ({\bf p}, \sigma) a ({\bf p}, \sigma) e^{-ip\cdot x} +
(\epsilon^\mu ({\bf p}, \sigma))^c b^\dagger ({\bf p}, \sigma) e^{+ip\cdot
x} \right ]\quad, \label{fo}
\end{equation}
the normalization of the wave
functions in the momentum representation is thus chosen to the unit,
$\epsilon_\mu^\ast ({\bf p}, \sigma) \epsilon^\mu ({\bf p},\sigma) = -
1$.\footnote{The metric used in this paper $g^{\mu\nu} = \mbox{diag} (1,
-1, -1, -1)$ is different from that of ref.~\cite{Wein}.} We observe that
in the massless limit all the defined polarization vectors of the momentum
space do not have good behaviour; the functions describing spin-1
particles tend to infinity.  This is not satisfactory.  Nevertheless,
after renormalizing the potentials, {\it e.~g.}, $\epsilon^\mu \rightarrow
u^\mu \equiv m \epsilon^\mu$ we come to the wave functions in the momentum
representation:\footnote{It is interesting to note that all the vectors
$u^\mu$ satisfy the condition $p_\mu u^\mu ({\bf p}, \sigma) = 0$.
It is relevant to the case of the Lorentz gauge and, perhaps, to the
analyses of the neutrino theories of light.}
\begin{mathletters} \begin{eqnarray} u^\mu
({\bf p}, +1)= -{N\over \sqrt{2}m}\pmatrix{p_r\cr m+ {p_1 p_r \over
E_p+m}\cr im +{p_2 p_r \over E_p+m}\cr {p_3 p_r \over
E_p+m}\cr}&\quad&,\quad u^\mu ({\bf p}, -1)= {N\over
\sqrt{2}m}\pmatrix{p_l\cr m+ {p_1 p_l \over E_p+m}\cr -im +{p_2 p_l \over
E_p+m}\cr {p_3 p_l \over E_p+m}\cr}\quad,\quad\label{vp12}\\ u^\mu ({\bf
p}, 0) &=& {N\over m}\pmatrix{p_3\cr {p_1 p_3 \over E_p+m}\cr {p_2 p_3
\over E_p+m}\cr m + {p_3^2 \over E_p+m}\cr}\quad, \quad  \label{vp3}
\end{eqnarray}
\end{mathletters}
($N=m$ and $p_{r,l} = p_1 \pm ip_2$) which do not
diverge in the massless limit.  Two of the massless functions (with
 $\sigma = \pm 1$) are equal to zero when the particle, described by this
field, is moving along the third axis ($p_1 = p_2 =0$,\, $p_3 \neq 0$).
The third one ($\sigma = 0$) is \begin{equation} u^\mu (p_3, 0)
\mid_{m\rightarrow 0} = \pmatrix{p_3\cr 0\cr 0\cr {p_3^2 \over E_p}\cr}
\equiv  \pmatrix{E_p \cr 0 \cr 0 \cr E_p\cr}\quad, \end{equation} and at
the rest ($E_p=p_3 \rightarrow 0$) also vanishes. Thus, such a field
operator describes the ``longitudinal photons" what is in the complete
accordance with the Weinberg theorem $B-A= h$ (let us remind  that we use
the $D(1/2,1/2)$ representation). Thus, the change of the normalization
can lead to the ``change" of physical content described by the classical
field (at least, comparing with the well-accepted one).  Of course, in the
quantum case one should somehow fix the form of commutation relations by
some physical principles.\footnote{I am {\it very} grateful to the
anonymous referee of my previous papers (``Foundation of Physics") who
suggested to fix them by requirements of the dimensionless of the action
(apart from the requirements of the translational and rotational
invariancies).} In the connection with the above consideration it is
interesting to remind that the authors of ref.~\cite[p.  136]{Itzyk} tried
to inforce the Stueckelberg's Lagrangian in order to overcome the
difficulties related with the $m\rightarrow 0$ limit (or the Proca theory
$\rightarrow$ Quantum Electrodynamics).  The Stueckelberg's Lagrangian is
well known to contain the additional term which may be put in
correspondence to some scalar (longitudinal) field (cf.
also~\cite{Staruz}).

Furthermore, it is easy to prove that the physical
fields $F^{\mu\nu}$ (defined by (\ref{1},\ref{2})) vanish in the
massless zero-momentum limit under the both definitions of normalization
and field equations. It is straightforward to find ${\bf
B}^{(+)} ({\bf p}, \sigma ) = {i\over 2m} {\bf p} \times {\bf u}
({\bf p}, \sigma)$\, , \, ${\bf E}^{(+)} ({\bf p}, \sigma) = {i\over 2m}
p_0 {\bf u} ({\bf p}, \sigma) - {i\over 2m} {\bf p} u^0
({\bf p}, \sigma)$ and the corresponding negative-energy strengths. Here
they are:\footnote{We assume that $[\epsilon^\mu ({\bf p},\sigma) ]^c
=e^{i\alpha_\sigma} [\epsilon^\mu ({\bf p},\sigma ) ]^\ast$, with
$\alpha_\sigma$ being arbitrary phase factors at this stage.
Thus, ${\cal C} = \openone_{4\times 4}$.
It is interesting to investigate other choices of the ${\cal C}$,
the charge conjugation matrix.}
\begin{mathletters} \begin{eqnarray}
{\bf B}^{(+)} ({\bf p},
+1) &=& -{iN\over 2\sqrt{2}m} \pmatrix{-ip_3 \cr p_3 \cr ip_r\cr} =
+ e^{-i\alpha_{-1}} {\bf B}^{(-)} ({\bf p}, -1 ) \quad,\quad   \label{bp}\\
{\bf B}^{(+)} ({\bf
p}, 0) &=& i{N \over 2m} \pmatrix{p_2 \cr -p_1 \cr 0\cr} =
- e^{-i\alpha_0} {\bf B}^{(-)} ({\bf p}, 0) \quad,\quad \label{bn}\\
{\bf B}^{(+)} ({\bf p}, -1)
&=& {iN \over 2\sqrt{2} m} \pmatrix{ip_3 \cr p_3 \cr -ip_l\cr} =
+ e^{-i\alpha_{+1}} {\bf B}^{(-)} ({\bf p}, +1)
\quad,\quad\label{bm}
\end{eqnarray}
\end{mathletters}
and
\begin{mathletters}
\begin{eqnarray}
{\bf E}^{(+)} ({\bf p}, +1) &=&  -{iN\over 2\sqrt{2}m} \pmatrix{E_p- {p_1
p_r \over E_p +m}\cr iE_p -{p_2 p_r \over E_p+m}\cr -{p_3 p_r \over
E+m}\cr} = + e^{-i\alpha^\prime_{-1}}
{\bf E}^{(-)} ({\bf p}, -1) \quad,\quad\label{ep}\\
{\bf E}^{(+)} ({\bf p}, 0) &=&  i{N\over 2m} \pmatrix{- {p_1 p_3
\over E_p+m}\cr -{p_2 p_3 \over E_p+m}\cr E_p-{p_3^2 \over
E_p+m}\cr} = - e^{-i\alpha^\prime_0} {\bf E}^{(-)} ({\bf p}, 0)
\quad,\quad\label{en}\\
{\bf E}^{(+)} ({\bf p}, -1) &=&  {iN\over
2\sqrt{2}m} \pmatrix{E_p- {p_1 p_l \over E_p+m}\cr -iE_p -{p_2 p_l \over
E_p+m}\cr -{p_3 p_l \over E_p+m}\cr} = + e^{-i\alpha^\prime_{+1}} {\bf
E}^{(-)} ({\bf p}, +1) \quad,\quad\label{em}
\end{eqnarray}
\end{mathletters}
where we denoted a normalization factor appearing in the
definitions of the potentials (and/or in the definitions of the physical
fields through potentials) as $N$. Let us note that as a result of the
above definitions we have
\begin{itemize}
\item
The cross products of   magnetic fields of different spin states (such as
${\bf B}^{(+)} ({\bf p}, \sigma) \times {\bf B}^{(-)} ({\bf p},
\sigma^\prime)$) may {\it not} be equal to zero and may be expressed  by
the ``time-like" potential (see the formula (\ref{tp})
below):\footnote{The relevant phase factors are assumed to be equal to zero.}
\begin{mathletters}
\begin{eqnarray}
{\bf B}^{(+)} ({\bf p}, +1) \times {\bf B}^{(-)} ({\bf p}, +1)
&=& - {iN^2 \over 4m^2} p_3 \pmatrix{p_1\cr p_2\cr
p_3\cr} = -{\bf B}^{(+)} ({\bf p}, -1) \times {\bf B}^{(-)} ({\bf p},
-1)\,\, ,\\
{\bf B}^{(+)} ({\bf p}, +1) \times {\bf B}^{(-)} ({\bf p}, 0)
&=& - {iN^2 \over 4m^2} {p_r \over \sqrt{2}} \pmatrix{p_1\cr
p_2\cr p_3\cr} = + {\bf B}^{(+)} ({\bf p}, 0) \times {\bf B}^{(-)} ({\bf
p}, -1)\,\, ,\\
{\bf B}^{(+)} ({\bf p}, -1) \times {\bf B}^{(-)} ({\bf p}, 0)
&=& - {iN^2 \over 4m^2} {p_l \over \sqrt{2}} \pmatrix{p_1\cr
p_2\cr p_3\cr} = + {\bf B}^{(+)} ({\bf p}, 0) \times {\bf B}^{(-)} ({\bf
p}, +1)\,\, .
\end{eqnarray}
\end{mathletters}
Other cross products are equal to zero.

\item
Furthermore, one can find the interesting relation:
\begin{eqnarray}
&&{\bf B}^{(+)} ({\bf p}, +1) \cdot {\bf B}^{(-)} ({\bf p},
+1) + {\bf B}^{(+)} ({\bf p}, -1) \cdot {\bf B}^{(-)} ({\bf p}, -1) + {\bf
B}^{(+)} ({\bf p}, 0) \cdot {\bf B}^{(-)} ({\bf p}, 0) =\nonumber\\
&&\qquad\qquad ={N\over 2m^2} (E_p^2 - m^2 )\, ,
\end{eqnarray}
due to
\begin{mathletters}
\begin{eqnarray}
&&{\bf B}^{(+)} ({\bf p}, +1) \cdot {\bf B}^{(-)} ({\bf p}, +1) =
{N^2 \over 8m^2} (p_r p_l +2p_3^2) = + {\bf B}^{(+)} ({\bf p}, -1)
\cdot {\bf B}^{(-)} ({\bf p}, -1)\,\, ,\\
&&{\bf B}^{(+)} ({\bf p}, +1) \cdot {\bf B}^{(-)} ({\bf p}, 0) =
{N^2 \over 4\sqrt{2} m^2} p_3 p_r = - {\bf B}^{(+)} ({\bf p}, 0)
\cdot {\bf B}^{(-)} ({\bf p}, -1)\,\, ,\\
&&{\bf B}^{(+)} ({\bf p}, -1) \cdot {\bf B}^{(-)} ({\bf p}, 0) =
-{N^2 \over 4\sqrt{2} m^2} p_3 p_l = - {\bf B}^{(+)} ({\bf p}, 0)
\cdot {\bf B}^{(-)} ({\bf p}, +1)\,\, ,\\
&&{\bf B}^{(+)} ({\bf p}, +1) \cdot {\bf B}^{(-)} ({\bf p}, -1) =
{N^2 \over 8 m^2} p_r^2\,\, ,\\
&&{\bf B}^{(+)} ({\bf p}, -1) \cdot {\bf B}^{(-)} ({\bf p}, +1) =
{N^2 \over 8 m^2} p_l^2\,\, ,\\
&&{\bf B}^{(+)} ({\bf p}, 0) \cdot {\bf B}^{(-)} ({\bf p}, 0) =
{N^2 \over 4 m^2} p_r p_l\,\, .
\end{eqnarray}
\end{mathletters}

\end{itemize}

For the sake of completeness let us
present the fields corresponding to the ``time-like" polarization:
\begin{equation}
u^\mu ({\bf p}, 0_t) = {N \over m} \pmatrix{E_p\cr p_1
\cr p_2\cr p_3\cr}\quad,\quad {\bf B}^{(\pm)} ({\bf p}, 0_t) = {\bf
0}\quad,\quad {\bf E}^{(\pm)} ({\bf p}, 0_t) = {\bf 0}\,\,.
\label{tp}
\end{equation}
The polarization vector $\epsilon^\mu ({\bf p}, 0_t)$ has
the good behaviour in $m\rightarrow 0$, $N=m$ (and also in the subsequent
limit ${\bf p} \rightarrow {\bf 0}$) and it may correspond to some
quantized field (particle).  Furthermore, in the case of the normalization
of potentials to the mass $N=m$  the physical fields which correspond to
the ``time-like" polarization  are equal to zero
identically.  The longitudinal fields
(strengths) are equal to zero in this limit only when one chooses the
frame  with $p_3 = \mid {\bf p} \mid$, cf. with the light front
formulation, ref.~\cite{DVALF}.  In the case $N=1$
and (\ref{1},\ref{2}) we have, in general,
the divergent behaviour of potentials and strengths.\footnote{In the case
of $N=1$ the fields ${\bf B}^\pm ({\bf p}, 0_t)$ and ${\bf E}^\pm ({\bf
p}, 0_t)$ would be undefined. This fact was also not appreciated in
the previous formulations of the theory of $(1,0)\oplus(0,1)$ and
$(1/2,1/2)$ fields.}

The spin operator recasts in the terms of the vector potentials as follows
(if one takes into account the dynamical equations, Eqs.
(\ref{de1},\ref{de2},\ref{1},\ref{2}))\footnote{The formula (\ref{spinf})
has certain similarities with the formula for the spin vector obtained
from Eqs.  (5.15,5.21) of ref.~\cite{Bogol}:
\begin{mathletters}
\begin{eqnarray}
{\bf J}_i &=& \epsilon_{ijk} \int J_{jk}^0 d^3 {\bf
x}\quad,\\ J_{\alpha\beta}^0 &=& \left ( A_\beta {\partial A_\alpha \over
\partial x_0} - A_\alpha {\partial A_\beta \over \partial x_0} \right
)\quad.
\end{eqnarray} \end{mathletters}
It describes the ``transversal photons" in the ordinary wisdom.
But, not all the questions related with the second $B_\mu$ potential,
the dual tensor $\widetilde{F}^{\mu\nu}$
and the normalization of the 4-potentials
have been clarified in the standard textbooks.}
\begin{eqnarray}
{\bf J}^k &=&  \epsilon^{ijk} \int d^3 {\bf x} \left [
F^{0i} (\partial_\mu F^{\mu j} ) + \widetilde F^{0i} (\partial_\mu
\widetilde F^{\mu j}) \right ] = \nonumber\\
&\qquad&\qquad = {1\over 4} \epsilon^{ijk} \int d^3 x \left [ B^j (
\partial^0 B^i - \partial^i B^0 ) - A^j ( \partial^0 A^i -\partial^i A^0 )
\right ] \quad.  \label{spinf}
\end{eqnarray}
If we put, as usual,
$\widetilde F^{\mu\nu} = \pm i F^{\mu\nu}$ (or $B^\mu =\pm A^\mu$) for the
right-  and left- circularly polarized radiation we would
obtain equating the spin operator to zero. The same situation would
occur if one chooses unappropriate normalization and/or if one uses the
equations (\ref{3},\ref{4}) without needed precautivity.  The
straightforward application of (\ref{3},\ref{4}) would lead
to the proportionality $J_{\kappa\tau} \sim m^2$ and, thus, it
appears that the spin operator would be equal to zero in the massless
limit, provided that the components of $A_\mu$ have good behaviour (do not
diverge in $m\rightarrow 0$).  Probably, this fact (the relation between
generators and the normalization)  was the origin of why many respectable
persons claimed that the antisymmetric tensor field would be pure
longitudinal. On the other hand, in the private communication Prof. Y.  S.
Kim stressed that neither he nor E. Wigner used the normalization of the
spin generators to the mass.  What is the situation which is realised in
the Nature (or both)? The answer still  depends on the choice of the field
operators, namely on the choice of positive- and negative- energy
solutions, creation/annihilation  operators and the normalization.

We note that not all the obscurities were clarified even in
our recent work~\cite{DVO96}.\footnote{First of all, we note that the
equality of the angular momentum generators to zero can be re-interpreted
as $$W_\mu P^\mu =0\,\, ,$$ with $W_\mu$ being the Pauli-Lubanski
operator.  This yields $$W_\mu = \lambda P_\mu\,\, ,$$ i.~e., what we need
in the massless case. But, according to the analysis above the 4-vector
$W_\mu$ would be equal to zero {\it identically} in the massless limit.
This is not satisfactory from the conceptual viewpoints.}
Let us calculate in a straightforward manner
the operator (\ref{spinf}).  If one uses the following definitions of
positive- and negative-energy parts of the antisymmetric tensor field
in the momentum space, i.~e., according to (\ref{bp}-\ref{em})
with ($\alpha_\sigma =0$):
\begin{equation}
(F_{\mu\nu})^{(+)}_{+1} = + (F_{\mu\nu})^{(-)}_{-1} \, , \,
(F_{\mu\nu})^{(+)}_{-1} = + (F_{\mu\nu})^{(-)}_{+1} \, , \,
(F_{\mu\nu})^{(+)}_{0} =  - (F_{\mu\nu})^{(-)}_{0} \quad.
\end{equation}
for the field operator
\begin{equation}
F_{\mu\nu} (x^\mu) = \int {d^3 {\bf p} \over (2\pi)^3 2E_p}
\sum_\sigma \left [
(F_{\mu\nu})^{(+)}_\sigma  ({\bf p}) a_\sigma ({\bf p}) e^{-ip\cdot x} +
(F_{\mu\nu})^{(-)}_\sigma  ({\bf p}) b_\sigma^\dagger ({\bf p})
e^{+ip\cdot x} \right ]\, ,
\end{equation}
then one obtains in the frame where $p_{1,2} =0$:
\begin{eqnarray}
{\bf J} &\equiv& {m\over 2}\int d^3 {\bf x}
{\bf E}(x^\mu) \times {\bf A} (x^\mu) =
{m^2 \over 4} \int {d^3 {\bf p} \over (2\pi)^3 \, 4E_p^2}
\left \{ \pmatrix{0\cr 0\cr E_p\cr} \right . \label{fin} \\
&&\left .\left [ a ({\bf p}, +1) b^\dagger ({\bf p},
+1) - a ({\bf p}, -1) b^\dagger ({\bf p}, -1) + b^\dagger ({\bf p}, +1) a
({\bf p}, +1) - b^\dagger ({\bf p}, -1) a ({\bf p}, -1)\right
] +\right .\nonumber\\
&+& \left. {E_p \over m \sqrt{2}} \pmatrix{E_p\cr iE_p \cr 0\cr}
\left [ a ({\bf p},+1) b^\dagger ({\bf p}, 0) +
b^\dagger ({\bf p}, -1) a ({\bf p}, 0)\right ] +\right .\nonumber\\
&+&\left . {E_p \over m \sqrt{2}} \pmatrix{E_p \cr -iE_p \cr 0\cr}
\left [ a ({\bf p}, -1) b^\dagger ({\bf p}, 0) +
b^\dagger ({\bf p}, +1) a ({\bf p}, 0) \right ] +\right .\nonumber\\
&+&\left . {1\over \sqrt{2}} \pmatrix{m\cr -im\cr 0\cr}
\left [ a ({\bf p}, 0) b^\dagger ({\bf p}, +1) +
b^\dagger ({\bf p}, 0)  a ({\bf p}, -1) \right ] +\right .\nonumber\\
&+&\left . {1\over \sqrt{2}} \pmatrix{m\cr im\cr 0\cr}
\left [ a ({\bf p}, 0) b^\dagger ({\bf p}, -1) +
b^\dagger ({\bf p}, 0) a ({\bf p}, +1) \right ]
\right \}\,\, .\nonumber
\end{eqnarray}
If the commutators of the states with
$h=\pm 1$ and $h=0$ are equal to zero
we have the ${\bf J}_3$ generator to be {\it non}-zero solely.
Above we used the fact that
\begin{mathletters}
\begin{eqnarray}
&& [(\partial_\mu F^{\mu j} ({\bf p},\sigma )]^{(+)}
= -{m\over 2} u^j ({\bf p},\sigma)\,\, ,\quad
[(\partial_\mu F^{\mu j} ({\bf p},\sigma )]^{(-)}
= -{m\over 2} [u^j ({\bf p},\sigma)]^\ast \,\, ,\\
&& [\partial_\mu \widetilde F^{\mu j} ({\bf p},\sigma) ]^\pm =0\,\, .
\end{eqnarray}
\end{mathletters}
The origin of this asymmetry can be discussed on the following basis:
while both $F^{\mu\nu}$ and $\widetilde F^{\mu\nu}$ can be expanded
in the potentials (cf. Eqs. (9) and (10)), but once choosing potentials in
order to obtain the fields (either $F^{\mu\nu}$ or its dual) we seem to be
no able to recover the former using the formulas relating $\widetilde
F^{\mu\nu} = {1\over 2} \epsilon^{\mu\nu\alpha\beta} F_{\alpha\beta}$.
All this might be related with the problems of the normalization
($N/m$) too. The  formulas (17,18) can describe both strengths ${\bf B}$
and ${\bf E}$, respectively, or vice versa. This conclusion is in the
complete accordance with the fact that for rigorous definition of the
parity one must use explicitly the representation $(j,0)\oplus (0,j)$, and
without proper definition of the creation/annihilation operators we are
not able to answer rigorously, which properties do ${\bf B}$ and ${\bf
E}$ have in the {\it momentum space} with respect to the space inversion
operation.

Next, it is obvious that though $\partial_\mu F^{\mu\nu}$
may be equal to zero in the massless limit from the formal viewpoint
and the equation (\ref{fin}) is proportional to the squared mass (?) for
the first sight, it is {\it not permitted} to forget that the commutation
relations may provide additional mass factors in the denominator of
(\ref{fin}).  It is the factor $\sim E_p/m^{2}$ in the commutation
relations\footnote{Remember that the dimension of the $\delta$ function is
inverse to its argument.}
\begin{equation} [a_\sigma ({\bf p}) ,
b^\dagger_{\sigma^\prime} ({\bf k}) ] \sim (2\pi)^3 {E_p\over m^2}
\delta_{\sigma\sigma^\prime} \delta ({\bf p} - {\bf k}) \,\, .
\end{equation}
which is required by the
principles of the rotational and translational
invariance\footnote{That is to say: the factor $\sim {1\over m^2}$
is required if one wants to obtain non-zero energy (and, hence,
helicity) excitations.} (and
also by the necessity of the description of the Coulomb long-range force
$\sim 1/r^2$ by means of the antisymmetric tensor field of the second
rank).

The dimension of the creation/annihilation operators of the 4-vector
potential should be [energy]$^{-2}$ provided that we use
(\ref{vp12},\ref{vp3}) with $N=m$.  Next, if we want the $F^{\mu\nu}
(x^\mu)$ to have the dimension [energy]$^2$ in the unit system $c=\hbar
=1$,\footnote{The dimensions [energy]$^{+1}$ of the field operators for
strengths was used here and in my previous paper in order  to keep
similarities with the Dirac case (the $(1/2,0)\oplus (0,1/2)$
representation) where the mass term presents explicitly in the term of the
bilinear combination of the fields.} we must divide the
Lagrangian\footnote{See the equation (3) of ref.~\cite{DVO96}, which was
also used in the present paper.} by $m^2$ (with the same $m$, the mass of
the particle!). This procedure will have the influence on the form of
(\ref{spinf},\ref{fin}) because the derivatives in this case pick up the
additional mass factor.  Thus, we can remove the ``ghost" proportionality
of the $c$- number coefficients in (\ref{fin}) to $\sim m^2$. The
commutation relations also change its form. The possibility of the above
renormalizations was {\it not} noted in the previous papers on the theory
of the 4-vector potenatial and  of the antisymmetric tensor field of the
second rank. Probably, this  was the reason of why peoples were confused
after including the mass factor of the creation/annihilation operators in
the field functions of $(1/2,1/2)$ and/or $(1,0)\oplus (0,1)$
representations.

Finally, we showed that the interplay between definitions of
the field functions and commutation relations occurs, thus giving
the {\it non-zero} values of the angular momentum generators
in the $(1,0)\oplus (0,1)$ representation.

The conclusion of the ``transversality" (in the meaning of existence of
$h=\pm 1$) is in accordance with the conclusion of the Ohanian's
paper~\cite{Ohanian}, cf.  formula (7) there:\footnote{Remember that in
almost all papers the electric field is defined to be equal to ${\bf E}^i
= F^{i0} = \partial^i A^0 - \partial^0 A^i$, with the potentials being not
well-defined in the massless limit of the Proca theory.  Usually, the
divergent part of the potentials was referred to the gauge-dependent part.
Furthermore, the physical fields and potentials were considered
classically in the cited paper, so the integration over the 3-momenta (the
quantization inside a cube) was not implied, see the formula (5) there.
Please pay also attention to the complex conjugation operation on
the potentials in the Ohanian's formula.  The formula (7) of
ref.~\cite{Ohanian} is in the SI unit system and our arguments above are
similar in the physical content. We did not still exclude the possibility
of the mathematical framework, which is different from our
presentation, but the conclusions, in my opinion,
must be in the accordance with the Weinberg theorem.}
\begin{equation} {\bf J} = {1\over 2\mu_0 c^2} \int \Re e ({\bf
E} \times {\bf A}^\ast ) \, d^3 {\bf x} = \pm {1\over \mu_0 c^2} \int
\frac{\hat {\bf z} E_0^2}{\omega} d^3 {\bf x} \,\, ,\label{spin}
\end{equation} with the Weinberg theorem, also with known experiments and
with the sane sense. The question, whether the situation could be realized
when the spin of the antisymmetric tensor field would be equal to zero,
must be checked by additional experimental verifications.
We do not exclude this possibility, founding our viewpoint on the
papers~\cite{Hayashi,Avdeev,Kirch,Ahl,Evgrav,Chub}. Finally, one
should note that we agree with the author of the cited
work~\cite[Eq.(4)]{Ohanian} about the gauge {\it non}-invariance of the
division of the angular momentum of the electromagnetic field into the
``orbital" and ``spin" part (\ref{spin}).

\acknowledgements
As a matter of fact, this paper has been inspired by remarks of the
referee of IJMPA (1994) and by the papers of Prof. D. V. Ahluwalia.
I am obliged to Profs. A. E. Chubykalo, M. W. Evans, Y. S. Kim, A.
Lakhtakia, A.  F.  Pashkov and S. Roy for illuminating discussions.


I am grateful to Zacatecas University for a professorship.
This work has been supported in part by the Mexican Sistema
Nacional de Investigadores, the Programa de Apoyo a la Carrera Docente
and by the CONACyT, M\'exico under the research project 0270P-E.


\begin{references}

\bibitem{EV1} M. W. Evans and J.-P. Vigier, {\it Enigmatic Photon.} Vols.
1-3 (Kluwer Academic Publishers, Dordrecht, 1994-96), the third volume
with S.  Jeffers and S. Roy. I still want to note that the reader
should be cautious working with the books and papers of M. Evans
because, unfortunately, they contain calculational and conceptual
errors. Nevertheless, I can but mention some bright ideas in his works.

\bibitem{DVO96R} V. V. Dvoeglazov, {\it Weinberg Formalism and New
Looks at the Electromagnetic Theory.} Preprint EFUAZ FT-96-35, Zacatecas,
Oct. 1996. Invited review for {\it The Enigmatic Photon.} Vol.  IV,
Chapter 12 (Kluwer, 1997) and references there.

\bibitem{WI} E. Inonu and E. Wigner, Proc. Natl. Acad. Sci. (USA) {\bf 39}
(1953) 510.

\bibitem{Kim} D. Han, Y. S. Kim and D. Son, Phys. Lett. {\bf 131}B (1983)
327, see also Y. S. Kim, Int. J. Mod. Phys. A{\bf 12} (1997)
71.

\bibitem{DVO96} V. V. Dvoeglazov,  {\it About the Claimed `Longitudinal
Nature' of the Antisymmetric Tensor Field After Quantization.} Preprint
EFUAZ FT-95-16-REV (hep-th/9604148), Zacatecas, Jan. 1996.

\bibitem{Hayashi} V. I. Ogievetskii and I. V. Polubarinov, Sov. J. Nucl.
Phys. {\bf 4} (1967) 156; K. Hayashi, Phys. Lett. B{\bf 44} (1973) 497; M.
Kalb and P.  Ramond, Phys. Rev. D{\bf 9} (1974) 2273.

\bibitem{Avdeev} L. V. Avdeev and M. V. Chizhov, Phys. Lett. B{\bf 321}
(1994) 212; {\it A queer reduction of degrees of freedom.} Preprint
JINR E2-94-263 (hep-th/9407067), Dubna: JINR, 1994.

\bibitem{WEIN0} S. Weinberg, Phys. Rev. B{\bf 134} (1964) 882

\bibitem{EV2} M. W. Evans, Found. Phys. {\bf 24} (1994)  1671.

\bibitem{EV3} M. W. Evans, Physica A{\bf 214} (1995) 605; Apeiron {\bf 4}
(1997) 48.

\bibitem{BW} V. Bargmann and E. P. Wigner, Proc. Natl. Acad. Sci. (USA)
{\bf 34} (1948) 211.

\bibitem{Lurie} D. Luri\`e, Particles and Fields (Interscience Publishers,
1968).

\bibitem{Kirch} M. Kirchbach, Mod. Phys. Lett. A{\bf 12} (1997) 2373.

\bibitem{Itzyk} C. Itzykson and J.-B. Zuber,
{\it Quantum Field Theory.} (McGraw-Hill Book Co. New York, 1980).

\bibitem{DVA} D. V. Ahluwalia and D. J. Ernst, Int. J. Mod. Phys. E{\bf 2}
(1993) 397; D. V. Ahluwalia, M. B. Johnson and T. Goldman, Phys. Lett.
B{\bf 316} (1993) 102.

\bibitem{Wein} S. Weinberg, {\it The Quantum Theory of Fields. Vol. I.
Foundations.} (Cambridge University Press, 1995).

\bibitem{Novozh} Yu. V. Novozhilov, {\it Introduction to Elementary
Particle Theory.} (Pergamon Press, Oxford, 1975).

\bibitem{Staruz} A. Staruszkiewicz,  Acta Phys. Polon. B{\bf 13}
(1982) 617; ibid {\bf 14} (1983) 63.

\bibitem{DVALF} D. V. Ahluwalia and M. Sawicki, Phys. Rev. D{\bf 47}
(1993) 5161

\bibitem{Bogol} N. N. Bogoliubov and D. V. Shirkov, {\it Introduction to
the Theory of Quantized Fields.} (Moscow, Nauka, 1973).

\bibitem{Ohanian} H. C. Ohanian, Am. J. Phys. {\bf 54} (1986) 500.

\bibitem{Ahl} D. V. Ahluwalia and D. J. Ernst, Mod. Phys. Lett. A{\bf 7}
(1992) 1967.

\bibitem{Evgrav} M. W. Evans, Found. Phys. Lett. {\bf 9} (1996) 397.

\bibitem{Chub} A. E. Chubykalo and R. Smirnov-Rueda, Mod. Phys. Lett. A
{\bf 12} (1997) 1.

\end{references}
\end{document}